\documentclass[journal=jacsat,manuscript=article]{achemso}
\setkeys{acs}{maxauthors = 10000}  
\setkeys{acs}{articletitle = true} 

\usepackage[version=3]{mhchem} 
\usepackage{xcolor} 
\usepackage{amsfonts} 
\usepackage{graphicx}
\usepackage{booktabs}
\SectionNumbersOn
\usepackage[colorlinks=false,hidelinks]{hyperref}

\newcommand*{\blauw}[1]{#1}

\author{Michael S. Jones}
\affiliation[UChicago]{Pritzker School of Molecular Engineering, University of Chicago, Chicago, Illinois 60637, United States}
\alsoaffiliation[equal]{These authors contributed equally to this work.}
\author{Kirill Shmilovich}
\affiliation[UChicago]{Pritzker School of Molecular Engineering, University of Chicago, Chicago, Illinois 60637, United States}
\alsoaffiliation[equal]{These authors contributed equally to this work.}
\author{Andrew L. Ferguson}
\affiliation[UChicago]{Pritzker School of Molecular Engineering, University of Chicago, Chicago, Illinois 60637, United States}
\email{andrewferguson@uchicago.edu}

\title{DiAMoNDBack: Diffusion-denoising Autoregressive Model for Non-Deterministic Backmapping of C$\alpha$ Protein Traces}

\begin{document}


\clearpage
\newpage

\begin{abstract}
\noindent Coarse-grained molecular models of proteins permit access to length and time scales unattainable by all-atom models and the simulation of processes that occur on long-time scales such as aggregation and folding. The reduced resolution realizes computational accelerations but an atomistic representation can be vital for a complete understanding of mechanistic details. Backmapping is the process of restoring all-atom resolution to coarse-grained molecular models. In this work, we report DiAMoNDBack (Diffusion-denoising Autoregressive Model for Non-Deterministic Backmapping) as an autoregressive denoising diffusion probability model to restore all-atom details to coarse-grained protein representations retaining only C$\alpha$ coordinates. The autoregressive generation process proceeds from the protein N-terminus to C-terminus in a residue-by-residue fashion conditioned on the C$\alpha$ trace and previously backmapped backbone and side chain atoms within the local neighborhood. The local and autoregressive nature of our model makes it transferable between proteins. The stochastic nature of the denoising diffusion process means that the model generates a realistic ensemble of backbone and side chain all-atom configurations consistent with the coarse-grained C$\alpha$ trace. We train DiAMoNDBack over 65k+ structures from Protein Data Bank (PDB) and validate it in applications to a hold-out PDB test set, intrinsically-disordered protein structures from the Protein Ensemble Database (PED), molecular dynamics simulations of fast-folding mini-proteins from DE Shaw Research, and coarse-grained simulation data. We achieve state-of-the-art reconstruction performance in terms of correct bond formation, avoidance of side chain clashes, and diversity of the generated side chain configurational states. We make DiAMoNDBack model publicly available as a free and open source Python package.
\end{abstract}

\newpage

\newpage

\section{Introduction}
Coarse-grained molecular models of proteins can substantially reduce the cost of molecular dynamics simulations and permit access to time and length scales and direct simulation of long-time processes such as folding, aggregation, and self-assembly that are inaccessible to all-atom molecular simulations~\cite{clementi2008coarse, noid2013perspective, saunders2013coarse, kmiecik2016coarse, mohr2022data, shmilovich2020discovery, kim2008coarse, scott2008coarse,lequieu20191cpn}. Coarse-graining achieves these accelerations by selectively clumping groups of atoms into super-atoms, or coarse-grained beads and deriving an associated coarse-grained force field with which to propagate the system dynamics. Eliminating degrees of freedom in the coarse-grained representation leads to computational accelerations associated with the reduced cost of tracking fewer particles, the possibility of larger numerical integration time steps, and smoothing of the underlying free energy landscape that accelerates phase space exploration of the coarse-grained system~\cite{marrink2013perspective, marrink2007martini, fritz2011multiscale}. Applications of coarse-graining to biomolecular dynamics have a rich history commencing with the pioneering work of Levitt and Warshel~\cite{levitt1975computer} and have led to a plethora of modern-day coarse-grained force fields such as MARTINI~\cite{marrink2007martini, monticelli2008martini, souza2021martini}, SPICA~\cite{seo2018spica}, Rosetta~\cite{das2008macromolecular}, PACE~\cite{han2010pace}, CABS~\cite{kolinski2004protein}, AWSEM~\cite{davtyan2012awsem}, and Upside~\cite{jumper2018accurate}. In recent years, there has been an explosion of interest in machine-learned coarse-grained potentials~\cite{zhang2018deepcg, wang2019machine, wang2019coarse, husic2020coarse, chennakesavalu2023ensuring, majewski2022machine, ding2022contrastive, durumeric2023machine, kohler2023flow, arts2023two, kramer2023statistically} that can be constructed from all-atom simulation data in a bottom-up fashion by rigorous variational techniques such as force matching~\cite{izvekov2005multiscale, noid2008multiscale} or relative entropy minimization~\cite{shell2008relative}. 

The primary concession of coarse-graining is a loss of atomistic detail that can be important in many applications such as determining atomistic contacts in protein-protein or protein-ligand interactions~\cite{badaczewska2020computational} or in downstream \textit{ab initio} calculations that require atomistic detail to compute properties such as dipole moments or NMR spectra~\cite{mcquarrie1997physical}. Backmapping is the process of reintroducing the lost degrees of freedom into a coarse-grained representation. This procedure can be conceived as a super-resolution task going from a coarse-grained to an atomistic geometry. The intrinsic loss of resolution in constructing a coarse-grained model means that the backmapping operation is one-to-many, and a primary challenge for backmapping algorithms is the generation of one or more physically realistic atomistic configurations associated with a particular coarse-grained structure. Contemporary backmapping methods can typically be categorized into either rules-based or data-driven approaches. Rules-based approaches use heuristics to produce an initial guess for the atomistic structure that is then refined using geometry optimization and/or energy minimization. The initial structures can be generated by querying fragment libraries~\cite{heath2007coarse, hess2006long, peter2009multiscale}, using random arrangements~\cite{rzepiela2010reconstruction}, or geometrically-guided initialization~\cite{lombardi2016cg2aa,wassenaar2014going, gopal2010primo, brocos2012multiscale, machado2016sirah, rotkiewicz2008fast}. Subsequent structural refinement and/or energy minimization is often necessary as the initial structures generated with rules-based approaches introduce unphysical artifacts such as atomistic clashes and/or anomalous bonds and dihedrals~\cite{nicholson2020constructing}. The requirement to manually adjust each backmapped structure introduces significant computational cost while also inherently biasing the final atomistic structure towards the particular choice of minimization scheme~\cite{badaczewska2020computational}. Rules-based approaches also tend to be deterministic in the sense that a particular coarse-grained structure will yield a single all-atom backmapped configuration. This can be an undesirable property since they fail to capture the thermodynamic ensemble of atomistic structures faithful to a single coarse-grained representation.


Data-driven techniques seek to remedy shortcomings of rules-based approaches by training neural networks to produce atomic structures conditioned on the coarse-grained representation~\cite{stieffenhofer2021adversarial, stieffenhofer2020adversarial, li2020backmapping, an2020machine, wang2022generative, shmilovich2022temporally, yang2023chemically}. These methods can achieve higher throughput compared to rules-based approaches as the models are trained to produce better well-equilibrated geometries that do not require a second stage of refinement or energy minimization. The most successful data-driven approaches tend to employ generative models, such as Variational AutoEncoders (VAEs)~\cite{kingma2013auto} and Generative Adversarial Networks (GANs)~\cite{goodfellow2020generative}, that produce atomistic structures conditioned on the coarse-grained structure as a model input and can learn to produce a distribution of backmapped atomistic structures~\cite{stieffenhofer2021adversarial, stieffenhofer2020adversarial, li2020backmapping, wang2022generative, shmilovich2022temporally, yang2023chemically}. While many of these data-driven techniques have demonstrated good performance when applied to relatively small biomolecules such as alanine dipeptide and chignolin~\cite{li2020backmapping, wang2022generative, shmilovich2022temporally}, they typically require training of bespoke models using atomistic training data and are not transferable to other molecules outside of the training data. A lack of transferability strongly limits the broader applicability of a backmapping model since training costs for one molecule cannot be amortized over other systems, and models cannot be developed for systems for which only coarse-grained trajectories are accessible and atomistic training data is either unavailable or insufficient to train a robust model. Work by Stieffenhofer et al.\ demonstrated potential for a transferable model by training on small molecule data and applying to polymer systems with their monomer units corresponding to the small molecules~\cite{stieffenhofer2021adversarial, stieffenhofer2020adversarial}. More recently, Yang and G\'omez-Bombarelli present the first instance of a chemically transferable backmapping model designed to backmap C$\alpha$ traces into full-resolution atomistic protein structures using a VAE architecture operating in the internal coordinate representation (dihedrals, angles, bond-lengths) of the protein~\cite{yang2023chemically}. The authors train on data from the Protein Ensemble Database (PED)~\cite{lazar2021ped}, which largely represents intrinsically disordered (IDP) proteins, and held out four PED proteins for testing and evaluation. 

\begin{figure}[ht!]
\centering
\includegraphics[width=0.94\columnwidth]{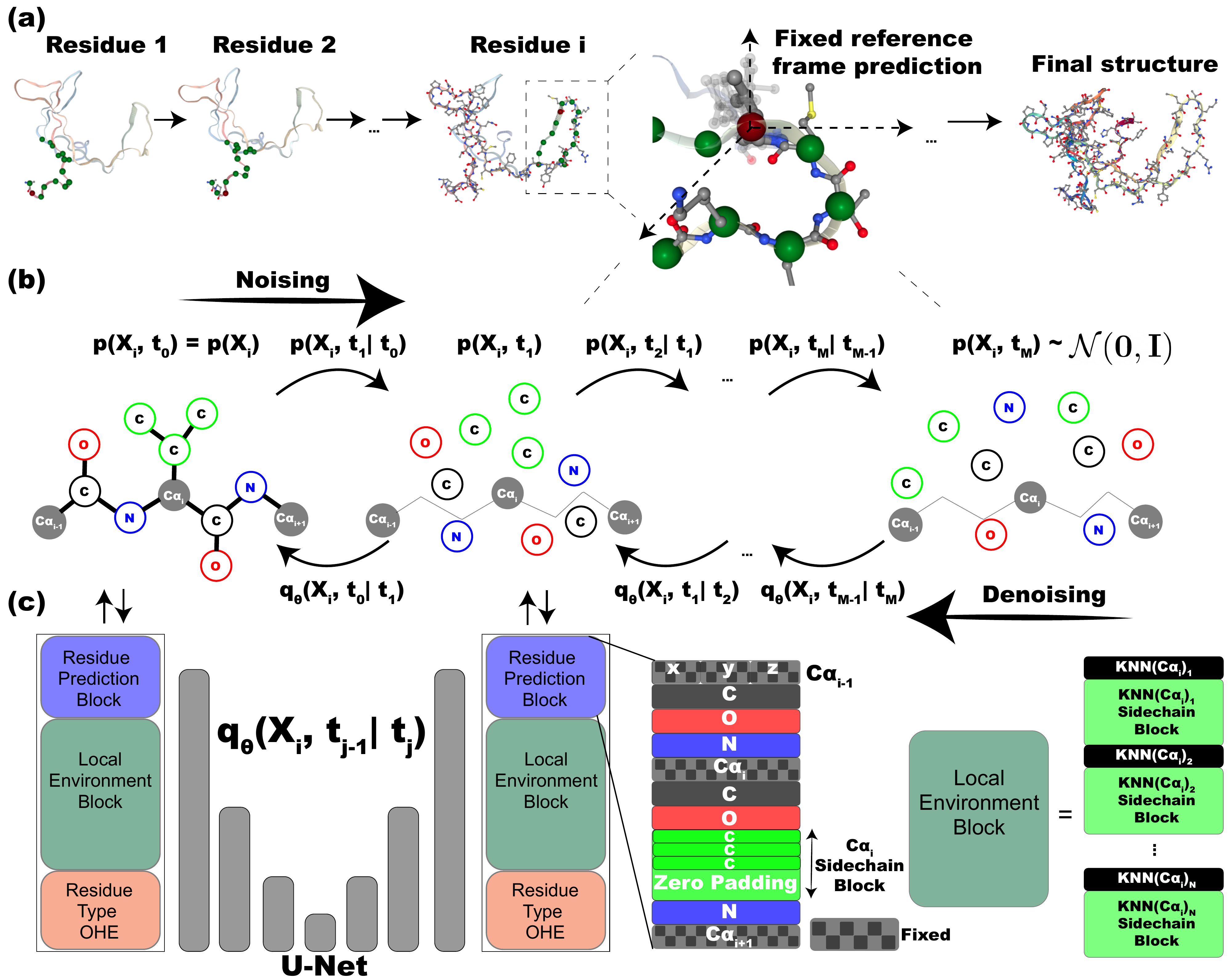}
\caption{Schematic illustration of DiAMoNDBack (Diffusion-denoising Autoregressive Model for Non-Deterministic Backmapping). \textbf{(a)} Atomistic detail is restored from C$\alpha$ traces by backmapping the structure residue-by-residue from the N-terminus to the C-terminus. Each residue prediction task is performed within a canonical reference frame and depends on the local environment around the target C$\alpha$ (red sphere) comprising the $N$ most spatially proximate residues to the target C$\alpha$ (green spheres). \textbf{(b)} The residue prediction task is performed using a Denoising Diffusion Probabilistic Model (DDPM) that learns to reverse an $M$-step noising process applied to real samples. The trained DDPM model can then operate on random noise to recover realistic-looking samples. \textbf{(c)} Learning the denoising process involves training a U-net that is designed to predict the noise added to a corrupted sample. The input to the network are Cartesian coordinate representations of the target residue, the local environment, and a one-hot encoding of the residue identity. Conditioning is achieved by only noising and regressing on components of the representation that are allowed to change throughout the diffusion steps in the residue prediction block such as the backbone C, N and O atoms and the side chain atoms. Atoms comprising the partially-decoded local environment and the C$\alpha$ backbone, along with the one-hot residue type encoding only serve as conditioning information passed to the network, are fixed throughout the diffusion steps, and do not contribute to the loss.} 
\label{fig:fig1}
\end{figure}

In this work, we present a transferable backmapping model for proteins termed DiAMoNDBack (Diffusion-denoising Autoregressive Model for Non-Deterministic Backmapping) (Fig.~\ref{fig:fig1}). The model is based on the recently popularized class of generative Denoising Diffusion Probabilistic Model (DDPM)~\cite{ho2020denoising, sohl2015deep}. DDPMs have demonstrated impressive performance within a number of molecular domains such as protein-ligand docking~\cite{corso2022diffdock,schneuing2022structure}, generation of molecular conformers~\cite{jing2022torsional, igashov2022equivariant}, learning of coarse-grained potentials~\cite{arts2023two}, and protein structure generation~\cite{wu2022protein, trippe2022diffusion, watson2022broadly, qiao2022dynamic, ingraham2022illuminating}. Our model is tasked to backmap atomistic proteins from C$\alpha$ traces by autoregressively generating atomistic structures in a residue-by-residue fashion from the N-terminus to C-terminus of the chain conditioned on the C$\alpha$ trace and any previously backmapped residues within the local neighborhood. The full protein structure is assembled by stitching together the backmapped residues along the coarse-grained C$\alpha$ backbone. Importantly, the local and autoregressive nature of our model makes it transferable between proteins, and the stochastic nature of the denoising diffusion process means that the model generates an ensemble of backbone and side chain configurations consistent with the coarse-grained C$\alpha$ trace. This means that we can both amortize the training cost of the model by applying it to arbitrary proteins outside of the training data, apply it to coarse-grained simulation trajectories for which no accompanying all-atom training data exists, and generate multiple physically-consistent realizations of the backbone and side chain configurations.

We train DiAMoNDBack over 65k+ structures from the curated from the ProteinNet~\cite{alquraishi2019proteinnet,king2021sidechainnet} database containing structures from the Protein Data Bank (PDB)~\cite{10.1093/nar/28.1.235, berman2003announcing} to construct a general-purpose generative model for backmapping protein structures from C$\alpha$ traces. We validate our model in applications to a hold-out PDB test set, and demonstrate the transferability of the model in applications to intrinsically-disordered protein structures from the Protein Ensemble Database (PED)~\cite{lazar2021ped}, molecular dynamics simulations of fast-folding mini-proteins from DE Shaw Research~\cite{lindorff2011fast}, and coarse-grained simulation data generated from bespoke coarse-grained potentials by Majewski et al.\ ~\cite{majewski2022machine}. We benchmark against the PULCHRA rules-based approach developed by Rotkiewicz and Skolnick~\cite{rotkiewicz2008fast} and the data-driven VAE-based GenZProt model developed by Yang and G\'omez-Bombarelli~\cite{yang2023chemically}. We achieve state-of-the-art reconstruction performance in terms of (i) correct formation of bonds, (ii) avoidance of side chain steric clashes, and (iii) diversity of the generated side chain configurational states. Contrary to the rules-based approach we do not suffer from a deterministic backmapping to a single structure~\cite{rotkiewicz2008fast}, and we better reproduce the natural distribution of side chain configurational states by avoiding the mode-collapse associated with the VAE-based approaches~\cite{yang2023chemically}. One drawback of the model is that the DDPM generation process is relatively slow, making it approximately 10$\times$ slower than GenZProt and 100$\times$ slower than PULCHRA. We make DiAMoNDBack model publicly available to the community as a free and open source Python package (see \blauw{Data Availability Statement}).

\section{Methods}

\subsection{Diffusion-denoising Autoregressive Model for Non-Deterministic Backmapping (DiAMoNDBack)}

Given a coarse-grained C$\alpha$ trace of a protein $\mathbf{x} = [x_1,  x_2,  \ldots,  x_{n-1},  x_{n}] \in \mathbb{R}^{n \times 3}$ with $x_{i} \in \mathbb{R}^{1 \times 3}$ representing the spatial localization of each of the $i = 1 \ldots n$ C$\alpha$ beads, the backmapping process seeks to learn the distribution of atomistic structures $p(\mathbf{X} \in \mathbb{R}^{N \times 3} | \mathbf{x})$ containing $N$ atoms consistent with and conditioned on the coarse-grained structure $\mathbf{x}$. We frame this reconstruction task in an autoregressive formulation~\cite{stieffenhofer2020adversarial,stieffenhofer2021adversarial,stieffenhofer2022benchmarking} where we generate the atomistic structure $\mathbf{X} = [X_1, X_2, \ldots, X_{n-1}, X_{n}]$ residue-by-residue, where $X_i \in \mathbb{R}^{p_i \times 3}$ represents the $p_i$ atoms associated with residue $i$ and $\sum_{i=1}^n p_i = N$. By decomposing the full distribution into a product of conditional distributions \\ $p(\mathbf{X} | \mathbf{x}) = p(X_1 | \mathbf{x})p(X_2 | \mathbf{x}, \{X_1\})p(X_3 | \mathbf{x}, \{X_1, X_2\})\ldots p(X_n | \mathbf{x}, \{X_1, X_2,\ldots,X_{n-2}, X_{n-1}\})$ we simplify the learning problem to backmapping residues $p(X_i | \mathbf{x}, \{X_1, X_2, \ldots, X_{i-2}, X_{i-1}\})$ in an autoregressive fashion rather than one-shot generation of the full protein structure (Fig.~\ref{fig:fig1}a). We solve this learning problem by training an autoregressive denoising diffusion probabilistic model (DDPM) \cite{ho2020denoising,sohl2015deep,luo2022understanding,yang2022diffusion,dhariwal2021diffusion} implemented within a conditional U-net architecture composed of 1D convolutional layers~\cite{ronneberger2015u, wang2022data} (Fig.~\ref{fig:fig1}c). Conditioning for each step of the residue-by-residue autoregressive backmapping is based on the C$\alpha$ trace, the $N$=14 most spatially proximate residues (i.e., have been brought into proximity by a secondary structural element, the tertiary fold, or quaternary complex, but which may be distantly separated in the backbone amino acid sequence), and a one-hot encoding of the residue type. Backmapping is performed by aligning each residue into a canonical alignment that permits us to directly generate the Cartesian coordinates of each backmapped atom in a rotationally and translationally invariant canonical reference frame that avoids the need for costly data augmentations otherwise required to implicitly learn insensitivity to rigid atomic rotations and translations~\cite{shmilovich2022temporally, shmilovich2022orbital}.

The DiAMoNDBack training protocol involves gathering an all-atom protein configuration from the training data, selecting a random residue index within the chain, selecting a random time step within the DDPM, sampling the commensurate degree of Gaussian noise to corrupt the sample, and training the U-net to predict the added noise and therefore learn to reverse the noising procedure (Fig.~\ref{fig:fig1}b). Once trained, the inference protocol generatively restores atomistic detail in a residue-wise fashion from the N to C-terminus of the C$\alpha$ representation of the protein chain. Specifically, we pass through each amino acid in an N-to-C fashion using the trained U-net to transform random Gaussian noise into coordinates of the constituent atoms of the residue. These coordinates are then used to update the chain representation that is used to condition backmapping of subsequent residues. The backmapping procedure for each residue therefore comprises four steps: (i) alignment into the canonical reference frame, (ii) featurization to extract the conditioning variables, (iii) inference of the predicted atomic coordinates via the DDPM implemented within the trained U-net, and (iv) realignment of the decoded residue into the protein backbone and incorporation of the atomic coordinates into the updated chain representation. Due to the challenges in representing terminal residues within a canonical reference frame~\cite{yang2023chemically}, N- and C-terminal residues are handled separately after first backmapping all of the internal residues. Multi-chain proteins are backmapped in the same order in which they appear in the structure file using the same intrachain N-to-C ordering. Inter-chain residues are treated analogously to intra-chain residues when constructing the local environment conditioning. As in the case of single-chain proteins, N- and C-termini are backmapped once all internal residues have been placed in all chains.

Full details of our mathematical formalism, DDPM loss function, conditional U-net architecture, residue featurization, canonical alignment process, treatment of N- and C-terminal residues, hyperparameter tuning -- including the choice of an N-to-C autoregressive ordering, use of a canonical Cartesian reference frame, and selection of $N$=14 most spatially proximate conditioning residues -- are provided in the \blauw{Supporting Information}.

\subsection{Data Curation}

We collated four data sets for DiAMoNDBack training and testing: (i) PED -- structural ensembles of primarily intrinsically disordered proteins~\cite{lazar2021ped}, (ii) PDB -- structures drawn from the RCSB Protein Data Bank~\cite{king2021sidechainnet,alquraishi2019proteinnet,10.1093/nar/28.1.235, berman2003announcing}, (iii) DES -- D.E.\ Shaw Research molecular dynamics simulations of fast folding mini proteins \cite{lindorff2011fast}, and (iv) CG -- coarse-grained simulations conducted by Majewski et al.~\cite{majewski2022machine}.

\textbf{PED.} For the purpose of comparison to the GenZProt model of Yang and G\'omez-Bombarelli~\cite{yang2023chemically} we train over the Protein Ensemble Database (PED)~\cite{lazar2021ped}, which contains structural ensembles of proteins including many intrinsically disordered proteins (IDPs). We discard three sequences -- PED00125e000, PED00126e000, and PED00161e002 -- that contain non-canonical amino acids, leaving us with 9228 structures comprising a total of 928,539 individual amino acid residue training samples. Following Yang and G\'omez-Bombarelli~\cite{yang2023chemically}, we adopted four PED proteins -- PED00151ecut0, PED00090e000, PED00055e000, and PED00218e000 -- containing 20-140 frames and including one two-chain protein (PED00218e000) as our test set, and employed the remaining data as out training set. Since GenZProt does not support backmapping of terminal residues, to make head-to-head comparisons with this model we report all quantitative analyses restricted to internal residues only.

\textbf{PDB.} Our primary production-level model was trained over protein structures collated from the Protein Data Bank (PDB)~\cite{10.1093/nar/28.1.235, berman2003announcing} held in the SidechainNet~\cite{king2021sidechainnet} extension of ProteinNet~\cite{alquraishi2019proteinnet} that itself builds on the data for the biennial Critical Assessment of protein Structure Prediction (CASP) challenges~\cite{moult2014critical}. For this PDB training data set we retain a majority of configurations but filter according to a number of criteria. We discarded any configuration that had four or more disconnected chains or contained a chain less than five residues long, leaving 98,665/103,716 sequences. Next, we removed any structures that include incomplete side-chain coordinates for any non-terminal residues resulting in the elimination of an additional 32,403 structures.
Finally, we eliminated 2,562 problematic structures containing one or more malformed neighboring C$\alpha$-C$\alpha$ distances lying outside the range of 2.7-4.1 \AA, where our cutoffs were informed by collating histograms of neighboring  C$\alpha$-C$\alpha$ distance distributions to identify outliers (\blauw{Fig.~S1}). This led us to retain a total 65,360 structures containing over 13M individual residue training samples (we note some structures were eliminated under multiple criteria and are not double-counted in the filtration). For the PDB test set, we employ the same test set as that provided by the ProteinNet database for the CASP12 blind structure prediction challenge~\cite{schaarschmidt2018assessment}. We filter the test set consistent with the criterion we used to filter the training data set removing structures that were missing some portion of the side chain atoms. After data cleaning, we extracted 24 test set proteins ranging in size from 60-599 residues that includes eight multi-chain proteins.

\textbf{DES.} The PED and PDB training data comprise static protein structures derived predominantly from experimental structure determination. These training examples are expected to largely correspond to structures lying in local or global minima of the configurational free energy landscape. We were interested to test if the performance of our model would improve with additional fine tuning on all-atom molecular dynamics (MD) trajectories containing a greater diversity of configurations including metastable states and transition states. We refined the model trained over the PDB training data by subjecting it to additional training over MD trajectories of 11 fast-folding mini proteins conducted by D.E.\ Shaw Research (DES)~\cite{lindorff2011fast}. We aggregated one complete trajectory of each protein, eliminating villin (2F4K) that contains a non-canonical amino acid residue, and strided each trajectory into 10,000 equally spaced frames. Using the procedure described in Sidky et al.~\cite{sidky2019high} we separated these frames into 100 contiguous chunks and randomly shuffled these chunks to form an 80/20 train/test split for each protein. In this way, the model is exposed to configurations across the full trajectory, but the test set retains regions that are temporally disjoint and distinct from the training data. To compile the fine-tuning data set we combined the training splits from each of the 11 proteins totaling 88,000 frames with an aggregate of 3,860,000 distinct residue training samples. When constructing our fine-tuning data set we find that performance for terminal residue prediction can substantially improve by over-representing terminal residue training examples by repeating their occurrences in the training data set (\blauw{Fig.~S11}). However, we find that there exists a trade-off where performance on internal residues begins to suffer if termini are too over-represented. For the fine-tuned models presented here, we employ a 5$\times$ augmentation of terminal residues, which we find to be a good balance in resolving terminal and internal residues with high fidelity. 


\textbf{CG.} Finally, we collected C$\alpha$ coarse-grained trajectories from the work of Majewski et al.~\cite{majewski2022machine} for three proteins of varying size: 1FME (28 residues), PRB (47 residues), and A3D (73 residues). In contrast to previous data sets, these are simulations carried out using a bespoke C$\alpha$-based coarse-grained force field. As such, there are no corresponding all-atom reference structures, so these data serve purely as testing data and a means to evaluate the out-of-domain generalization and transferability of our model. We performed even striding across all available 32 trajectories for each protein to obtain 2,000 frames for each system. 

\begin{figure}[ht!]
\centering
\includegraphics[width=\columnwidth]{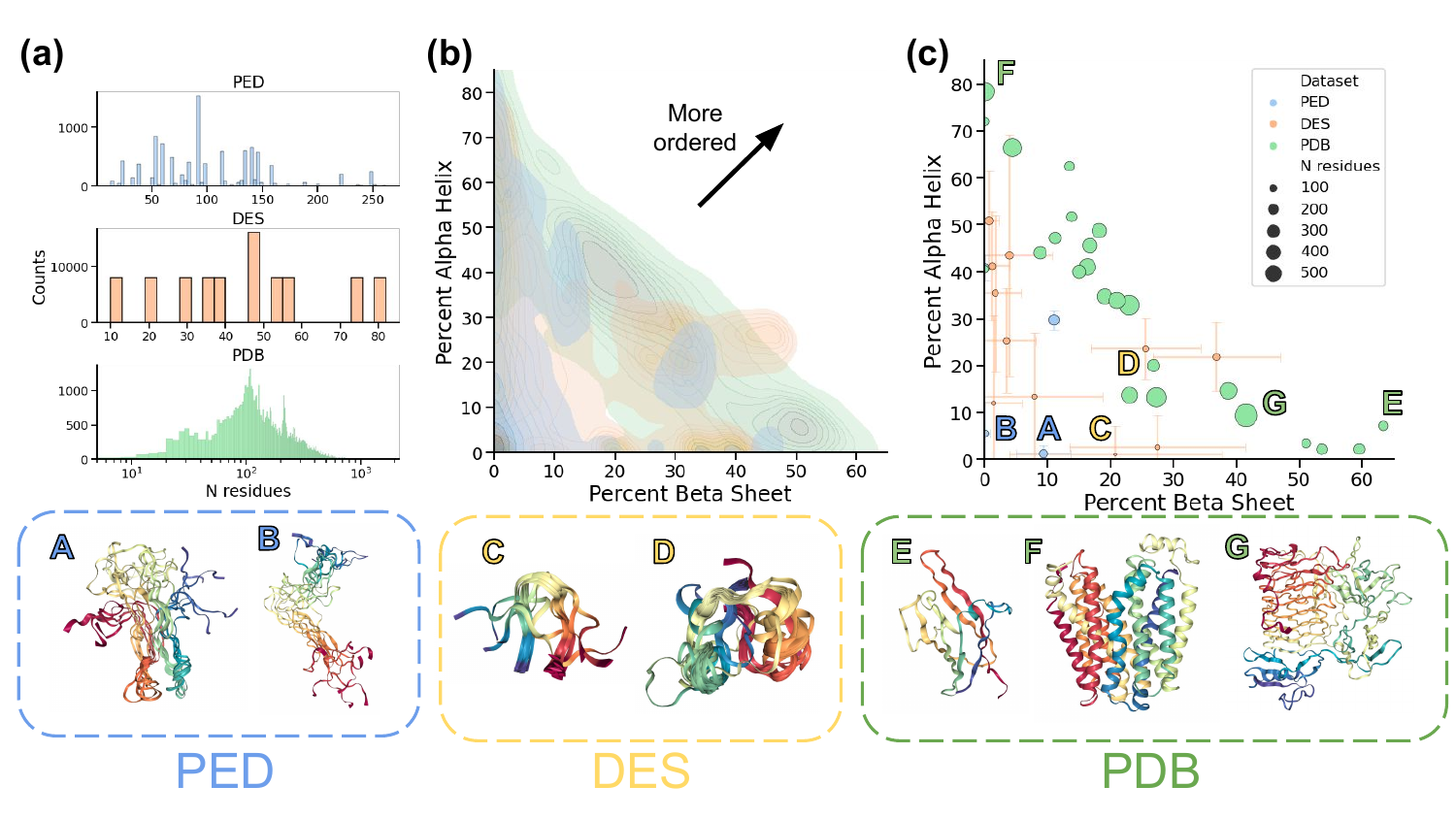}
\caption{Visual analysis of protein sequences used for training and testing in the PED, PDB, and DES training sets. \textbf{(a)} Distribution of sequence lengths. \textbf{(b)} Distribution of structural diversity in the space of average percent alpha-helix and beta-sheet content. Disordered/unstructured structures reside close to the origin. Colors correspond to the histograms in panel (a): PED = blue, PDB = green, and DES = orange. Darker regions represent areas with higher probability density. \textbf{(c)} Test set proteins for each training data set shown in the same space of alpha-helix and beta-sheet content. The size of each point corresponds to the number of residues in the protein. For the PED and DES data, multiple frames are present for each protein and error bars indicate the standard deviation across frames. Selected representative training structures are labeled A-F and the corresponding structures rendered along the bottom of the figure.} 
\label{fig:fig2}
\end{figure}

A visual comparison of the four data sets is presented in Fig.~\ref{fig:fig2}. In Fig.~\ref{fig:fig2}a we illustrate the distribution of sequence lengths. The PED training data comprises proteins of 13-260 residues in length with a total of 96 sequences and 9788 configurations. The DES data contains 11 sequences ranging from the small chignolin protein containing just 10 residues to the large $\lambda$-repressor containing 80 residues for a total of 88,000 structures. The PDB training data set contains the largest diversity of proteins of 5-2082 residues in length and comprises 65,360 structures. Alongside the sequence diversity, we represent the structural diversity of our training data sets by visualizing their distribution in the space of alpha-helical and beta-sheet content (Fig.~\ref{fig:fig2}b). The PED data tends toward relatively low alpha-helix and beta-sheet content, reflecting the intrinsically disordered nature of the data set. The PDB data spans a wide range of alpha-helix and beta-sheet content indicative of the more globular and ordered structures that originate from crystallographic data. The DES data, while containing the least sequence diversity, covers a wide range of structural diversity due to the sampling of both folded and unfolded configurations in the MD simulations. In Fig.~\ref{fig:fig2}c, we illustrate the hold out test set proteins corresponding to each of the three training data sets in the space of their alpha-helix and beta-sheet content along with visualizations of selected structures. In \blauw{Fig.~S5} we present a residue-level comparison of the representation of the 20 natural amino acids within the PED, PDB, and DES training data.

\subsection{Evaluation metrics} \label{sec:metrics}

We define three metrics to evaluate the quality of our backmapping predictions that measure (1) the quality of the reconstructed atomic bonds, (2) the degree of non-bonded steric clashes between residues, and (3) the diversity of atomistic structures produced. 

\textbf{Bond score ($\uparrow$).} The quality of the bond graph for our backmapped structures is determined by calculating the percentage of bonded atom distances that lie within 10\% of the bond distance in the reference atomistic structure. For the coarse-grained trajectories in the CG test data that do not possess corresponding atomistic reference trajectories, we use the average values for bond lengths from the test set DES data for the associated protein as the reference. The bond score varies between 0-100\% with higher values indicating better performance.

\textbf{Clash score ($\downarrow$).} A major failure mode for backmapping is the the construction of side-chain placements that result in unphysical steric clashes. We quantify the degree of steric clashes by computing the fraction of residues in which one or more atoms lie within 1.2 \AA\ of the atoms of another residue. For neighboring residues in the protein backbone, only clashes between the side chain atoms are considered. The 1.2 \AA\ threshold was selected based on Yang and G\'omez-Bombarelli~\cite {yang2023chemically} adopting this cutoff for defining atomic clashes. The clash score varies between 0-100\% with lower values indicating better performance.

\textbf{Diversity score ($\downarrow$).} A single coarse-grained C$\alpha$ trace is consistent with multiple all-atom configurations. A desirable trait of a backmapping model is its capacity to generate an ensemble of all-atom predictions each of which is compatible with the C$\alpha$ trace and itself contains well-formed bonds and avoids steric clashes. Previous work has used the root-mean squared distance (RMSD) between samples as a metric for generative diversity, where a high RMSD between samples indicates high diversity~\cite{wang2022generative}. However, a low RMSD between the samples and the atomistic reference has also been used to show good adherence to conditioning and faithful reconstruction~\cite{yang2023chemically}. We combine these two desiderata to posit that the single all-atom reference structure should be indistinguishable from the distribution of generated configurations. As a corollary, the average pairwise RMSDs between (i) all generated samples and the reference and (ii) all generated samples with themselves should be approximately equal. This leads us to define a generative diversity score $DIV$ as,
\begin{align} 
RMSD_{ref} &= \frac{1}{G}\sum_{i}^{G}{RMSD(\mathbf{X}_{i}^{gen}}, \mathbf{X}^{ref}) \label{eqn:ref} \\
RMSD_{gen} &= \frac{2}{G(G-1)}\sum_{i}^{G} \sum_{j}^{(i-1)}{RMSD(\mathbf{X}_{i}^{gen}, \mathbf{X}_{j}^{gen}}) \label{eqn:gen}\\
DIV &= 1 - \frac{RMSD_{gen}}{RMSD_{ref}} \label{eqn:div}
\end{align}
where $G$ is the number of generated samples $\{\mathbf{X}_{1}^{gen}, \ldots, \mathbf{X}_{G}^{gen}\}$ conditioned on the C$\alpha$ trace possessing a single reference configuration $\mathbf{X}^{ref}$. 

As shown in the \blauw{Supporting Information}, $RMSD_{gen} \approx \frac{2}{(G-1)\sqrt{G}}\sum_{i}^{G}{RMSD(\mathbf{X}_{i}^{gen}}, \overline{\mathbf{X}^{gen}})$, where $\overline{\mathbf{X}^{gen}} = \frac{1}{G}\sum_{i}^{G}{\mathbf{X}_{i}^{gen}}$ is the mean generated configuration and we assume the inequalities used in the derivation to be tight. Approximating $\frac{2}{(G-1)\sqrt{G}} \approx \frac{1}{G}$ allows us to interpret $RMSD_{gen}$ as the average RMSD of the generated configurations around their own mean. This construction is instructive as it allows us to then approximately interpret the diversity score $DIV$ as a comparison of the average RMSD of the generated configurations around the reference configuration $\mathbf{X}^{ref}$ relative to that around their own mean $\overline{\mathbf{X}^{gen}}$. In general, one would anticipate a tighter distribution around the mean of a distribution than any other imposed point, such that we would expect $RMSD_{gen}$$<$$RMSD_{ref}$. This, however, is not a strict inequality and one may expect it to be violated, particularly for small values of $G$ that produces noisy estimates of $\overline{\mathbf{X}^{gen}}$. Nevertheless, our empirical calculations show that in the present applications this inequality generally holds with only occasional violations, and that our calculated diversity scores approximately lie on the interval [0,1]. Diversity scores of unity are obtained for deterministic backmapping since $RMSD_{gen}$=0, while diversity scores close to zero are achieved for $RMSD_{ref}$$\approx$$RMSD_{gen}$ and are indicative of better model performance. 
We note that the diversity metric cannot be calculated for the coarse-grained data that lack reference atomistic structures, but we can still compute $RMSD_{gen}$ as a proxy metric for generative diversity.  

\section{Results and Discussion}

We train and evaluate DiAMoNDBack on four train/test data sets and benchmark its performance against GenZProt as a state-of-the-art the VAE-based deep generative model~\cite{yang2023chemically} and the PULCHRA rules-based approach that generates an approximate structure using heuristics and atomic fragments followed by rotating side chain dihedrals to resolve steric clashes~\cite{rotkiewicz2008fast}. We choose not perform any energy minimization or molecular mechanics relaxations for any of the three models to compare and evaluate the backmapping approaches independent of any molecular force-field and the high computational cost associated with these operations for large proteins. We also note that while we provide routines for handling prediction of terminal residues, for the purposes of comparison to GenZProt, which does not support termini prediction, all analyses herein are performed on protein structures stripped of terminal residues. Details of our usage and application of PULCHRA and GenZProt and an analysis of the quality of DiAMoNDBack terminal residue backmapping is provided in the \blauw{Supporting Information}.

\subsection{PED training and evaluation}

Following Yang and G\'omez-Bombarelli~\cite {yang2023chemically}, we first train DiAMoNDBack on the PED training data set comprising conformational ensembles of intrinsically disordered proteins~\cite{lazar2021ped}. A numerical comparison of DiAMoNDBack and GenZProt performance on the set of four held-out PED test examples is presented in Table~\ref{tab:PED}. In terms of bond score, performance of DiAMoNDBack and GenZProt are both excellent, lying above 95\% correctly formed bonds in all cases and with essentially indistinguishable performance. In terms of clash score, DiAMoNDBack outperforms GenZProt with an overall improvement of $\sim$50\% fewer clashing residues across the four structures and superior clash scores to GenZProt on all test proteins except PED00151ecut0 where our mean clash score is slightly better than GenZProt but not outside of error. In terms of diversity score, DiAMoNDBack shows a substantial improvement in generative diversity relative to GenZProt, achieving diversity scores of 0.23 or better on all four test examples while GenZProt scores are no better than 0.85. (We recall from Sec~\ref{sec:metrics} that lower diversity scores are indicative of superior model performance -- deterministic models with no conformational diversity possess diversity scores of unity, whereas models in which the average diversity between generated configurations matches that of the generated configurations with the reference have diversity scores of zero.) As discussed further in Sec.~\ref{sec:dih}, the improved conformational diversity of DiAMoNDBack relative to GenZProt appears to be at least partially attributable to the VAE-based model suffering from mode collapse and failing to generate a high diversity of side chain dihedral angles.


\begin{table}[]
\centering
\resizebox{\columnwidth}{!}{%
\begin{tabular}{@{}ccccc@{}}
\cmidrule(l){2-5}
           & \multicolumn{4}{c}{Test protein}                                                        \\ \cmidrule(l){2-5} 
           & PED00055e000      & PED00090e000            & PED00151ecut0           & PED00218e000    \\ \cmidrule(l){2-5} 
           & Bond ($\uparrow$) {[}\%{]}   & Bond ($\uparrow$) {[}\%{]}         & Bond ($\uparrow$) {[}\%{]}         & Bond ($\uparrow$) {[}\%{]} \\ \cmidrule(l){2-5} 
GenZProt   & 97.55$\pm$0.02    & 95.30$\pm$0.02          & \textbf{97.05$\pm$0.01} & 98.06$\pm$0.01  \\
DiAMoNDBack (PED) & 97.70$\pm$0.08    & \textbf{97.94$\pm$0.07} & 96.76$\pm$0.05          & 98.07$\pm$0.10  \\ \cmidrule(l){2-5} 
           & Clash ($\downarrow$) {[}\%{]}   & Clash ($\downarrow$) {[}\%{]}         & Clash ($\downarrow$) {[}\%{]}         & Clash ($\downarrow$) {[}\%{]} \\ \cmidrule(l){2-5} 
GenZProt   & 4.67$\pm$0.12     & 9.27$\pm$0.13           & 0.34$\pm$0.07           & 5.42$\pm$0.28   \\
DiAMoNDBack (PED) & \textbf{2.82$\pm$0.18}   & \textbf{5.04$\pm$0.68}   & 0.30$\pm$0.03            & \textbf{1.76$\pm$0.38}   \\ \cmidrule(l){2-5} 
           & Diversity ($\downarrow$)        & Diversity ($\downarrow$)             & Diversity ($\downarrow$)              & Diversity ($\downarrow$)     \\ \cmidrule(l){2-5} 
GenZProt   & 0.9048$\pm$0.0004 & 0.888$\pm$0.002         & 0.8533$\pm$0.0007       & 0.862$\pm$0.001 \\
DiAMoNDBack (PED) & \textbf{0.221$\pm$0.002} & \textbf{0.208$\pm$0.003} & \textbf{0.148$\pm$0.001} & \textbf{0.187$\pm$0.006}
\end{tabular}%
}
\caption{Comparison of DiAMoNDBack and GenZProt trained over the PED training data and evaluated on the four held-out PED test examples. Bond scores enumerate the fraction of correctly formed atomistic bond lengths with higher values on the 0-100\% range associated with superior performance. Clash scores enumerate the fraction of residues engaged in physically unrealistic steric clashes with lower values on the 0-100\% range associated with superior performance. Diversity scores measure the configurational diversity of the generated atomistic configurations and approximately lie on a [0,1] range. Lower diversity scores are associated with superior performance: deterministic models with no configurational diversity possess a diversity score of unity, whereas models producing an average diversity between generated configurations matching that of the generated configurations with the reference have diversity scores of zero. Standard deviations in the reported values are estimated using five-fold block averaging for the bond and clash scores and using jackknife resampling for the diversity scores. The model exhibiting superior performance in any category outside of error bars is indicated in \textbf{bold}.}
\label{tab:PED}
\end{table}

\subsection{PDB and DES training and evaluation}

The PED data set is both relatively small and the intrinsically disordered character of the constituent proteins means that models trained on these data are not representative of globular and folded structures typically associated with functional proteins. We therefore trained our production-level DiAMoNDBack model, termed DiAMoNDBack (PDB), over 65k+ structures collated from the Protein Data Bank (PDB)~\cite{king2021sidechainnet,alquraishi2019proteinnet,10.1093/nar/28.1.235, berman2003announcing} comprising 680$\times$ more sequences, 6.7$\times$ more configurations, and 15$\times$ more individual residue training examples than PED. The PDB structures predominantly reside in local or global minima of the configurational free energy landscape and may therefore underrepresent transient and metastable states. As such, we also fine-tuned our PDB trained model over the DES data set comprising long simulation trajectories of 11 fast-folding mini-proteins generated by D.E.\ Shaw Research~\cite{lindorff2011fast} to produce our fine-tuned production-level model DiAMoNDBack (PDB;DES-FT). We hypothesized that the fine-tuned model should also be better calibrated to the DES force field and represent an example of developing a bespoke force field-specific variant of the baseline DiAMoNDBack (PDB) model using modest amounts of all-atom simulation data.

In Table~\ref{tab:PDBDES} we present a comparison of the performance of PULCHRA~\cite{rotkiewicz2008fast}, GenZProt~\cite{yang2023chemically}, DiAMoNDBack (PDB), and DiAMoNDBack (PDB;DES-FT) on the 24 proteins comprising the held-out PDB test set and the held-out test split for the 11 DES all-atom simulation trajectories. We report the bond, clash, and diversity scores averaged over the test sets and also the best and worst performing systems as judged by the clash score to illustrate the performance range. 

\begin{table}[]
\resizebox{\columnwidth}{!}{%
\begin{tabular}{@{}ccccccc@{}}
\cmidrule(l){2-7}
 &
  \begin{tabular}[c]{@{}c@{}}PDB\\ overall\end{tabular} &
  \begin{tabular}[c]{@{}c@{}}PDB best\\ (TBM\#T0922)\end{tabular} &
  \begin{tabular}[c]{@{}c@{}}PDB worst\\ (TBM-hard\#T0912)\end{tabular} &
  \begin{tabular}[c]{@{}c@{}}DES\\ overall\end{tabular} &
  \begin{tabular}[c]{@{}c@{}}DES best\\ (NTL9)\end{tabular} &
  \begin{tabular}[c]{@{}c@{}}DES worst\\ (UVF)\end{tabular} \\ \cmidrule(l){2-7} 
 &
  Bond ($\uparrow$) {[}\%{]} &
  Bond ($\uparrow$) {[}\%{]} &
  Bond ($\uparrow$) {[}\%{]} &
  Bond ($\uparrow$) {[}\%{]} &
  Bond ($\uparrow$) {[}\%{]} &
  Bond ($\uparrow$) {[}\%{]} \\ \cmidrule(l){2-7} 
PULCHRA &
  98.91 &
  99.82 &
  98.41 &
  98.45 &
  98.7 &
  99.2 \\
GenZProt &
  96.253$\pm$0.005 &
  97.71$\pm$0.22 &
  94.20$\pm$0.024 &
  94.855$\pm$0.002 &
  96.41$\pm$0.003 &
  95.905$\pm$0.002 \\
DiAMoNDBack (PDB) &
  \textbf{99.18$\pm$0.04} &
  99.78$\pm$0.18 &
  98.97$\pm$0.09 &
  97.981$\pm$0.002 &
  98.24$\pm$0.01 &
  97.22$\pm$0.01 \\
DiAMoNDBack (PDB;DES-FT) &
  98.99$\pm$0.06 &
  99.56$\pm$0.09 &
  98.61$\pm$0.25 &
  \textbf{98.725$\pm$0.004} &
  98.88$\pm$0.01 &
  98.26$\pm$0.01 \\ \cmidrule(l){2-7} 
 &
  Clash ($\downarrow$) {[}\%{]} &
  Clash ($\downarrow$) {[}\%{]} &
  Clash ($\downarrow$) {[}\%{]} &
  Clash ($\downarrow$) {[}\%{]} &
  Clash ($\downarrow$) {[}\%{]} &
  Clash ($\downarrow$) {[}\%{]} \\ \cmidrule(l){2-7} 
PULCHRA &
  \textbf{0.15} &
  0 &
  1.01 &
  0.20 &
  0.26 &
  0.24 \\
GenZProt &
  8.43$\pm$0.22 &
  10.56$\pm$1.11 &
  14.81$\pm$0.43 &
  6.01$\pm$0.04 &
  1.43$\pm$0.01 &
  6.96$\pm$0.04 \\
DiAMoNDBack (PDB) &
  0.57$\pm$0.09 &
  0.00$\pm$0.00 &
  1.07$\pm$0.44 &
  0.33$\pm$0.01 &
  0.12$\pm$0.02 &
  0.58$\pm$0.04 \\
DiAMoNDBack (PDB;DES-FT) &
  0.52$\pm$0.16 &
  0.00$\pm$0.00 &
  0.91$\pm$0.47 &
  \textbf{0.175$\pm$0.007} &
  0.04$\pm$0.01 &
  0.38$\pm$0.02 \\ \cmidrule(l){2-7} 
 &
  Diversity ($\downarrow$) &
  Diversity ($\downarrow$) &
  Diversity ($\downarrow$) &
  Diversity ($\downarrow$) &
  Diversity ($\downarrow$) &
  Diversity ($\downarrow$) \\ \cmidrule(l){2-7} 
PULCHRA &
  1 &
  1 &
  1 &
  1 &
  1 &
  1 \\
GenZProt &
  0.865$\pm$0.002 &
  0.84$\pm$0.01 &
  0.884$\pm$0.001 &
  0.8316$\pm$0.0001 &
  0.8758$\pm$0.0001 &
  0.8962$\pm$0.0001 \\
DiAMoNDBack (PDB) &
  \textbf{0.037$\pm$0.004} &
  0.16$\pm$0.03 &
  -0.004$\pm$0.015 &
  0.0329$\pm$0.0002 &
  0.0706$\pm$0.0004 &
  0.0230$\pm$0.0002 \\
DiAMoNDBack (PDB;DES-FT) &
  0.064$\pm$0.004 &
  0.09$\pm$0.03 &
  0.076$\pm$0.008 &
  \textbf{0.0223$\pm$0.0002} &
  0.0235$\pm$0.0002 &
  0.0206$\pm$0.0005 \\ \cmidrule(l){2-7} 
\end{tabular}%
}
\caption{Comparison of backmapping performance on PDB data and MD trajectories between the rules-based PULCHRA approach~\cite{rotkiewicz2008fast}, GenZProt~\cite{yang2023chemically}, DiAMoNDBack trained on the PDB data set DiAMoNDBack (PDB), and DiAMoNDBack fine-tuned on MD trajectory data DiAMoNDBack (PDB;DES-FT). The ``PDB overall'' and ``DES overall'' columns report aggregate metrics averaged over all samples and frames. Two additional columns reports metrics on the best- and worst-performing systems as determined by clash score to give an appreciation for the performance range. Standard deviations in the reported values for DiAMoNDBack and GenZProt are estimated using five-fold block averaging for the bond and clash scores and using jackknife resampling for the diversity scores. As a deterministic algorithm, the values reported for PULCHRA do not have associated uncertainties and the diversity scores for this model are, by definition, unity. We note that overall metrics for PULCHRA on the PDB test set are only evaluated on the 16/24 single-chain proteins, as the software failed to operate on multi-chain systems. The model exhibiting superior performance in any category outside of error bars in the PDB overall and DES overall tasks is indicated in \textbf{bold}.}
\label{tab:PDBDES}
\end{table}

The bond scores averaged over the PDB and DES test sets are better than 94\% for all four models. GenZProt slightly underperforms the other three models by 2-3 percentage points, but the difference is rather small.

The clash scores expose a more significant performance spread among the models. For GenZProt, 8.43\% (PDB) and 6.01\% (DES) of residues are positioned in unphysical steric clashes when averaged over the test sets. DiAMoNDBack (PDB) performs more than an order of magnitude better with clash scores of 0.57\% (PDB) and 0.33\% (DES), and the fine-tuned DiAMoNDBack (PDB;DES-FT) model is better still at 0.52\% (PDB) and 0.18\% (DES). The rules-based PULCHRA model is almost as good as the fine-tuned DiAMoNDBack on the DES data at 0.20\% (DES) but is superior on the PDB at 0.15\% (PDB). The superior performance of DiAMoNDBack (PDB;DES-FT) relative to DiAMoNDBack (PDB) on the DES test set is expected, and illustrates the value of fine-tuning a bespoke model for a particular force field. The small performance improvement on the PDB test data, or at least the absence of any performance degradation within error bars, was more surprising and suggests that the fine-tuned model is not overfitting to the DES data and maintaining a transferable and generic backmapping model. The extremely good clash score performance of PULCHRA is unsurprising since this algorithm attempts to explicitly resolve steric clashes by rotating side chain dihedrals after placement of the residues. It is particularly encouraging, therefore, that DiAMoNDBack (PDB;DES-FT) is competitive with and/or superior to PULCHRA in this metric.

The diversity score of the deterministic rules-based PULCHRA model that generates a single backmapped configuration is, by definition, unity. GenZProt improves upon this slightly to achieve scores of 0.87 (PDB) and 0.83 (DES) but the proximity of these values to unity indicates that the preponderance of configurations are structurally very similar and the configurational diversity one would expect to be present within the ensemble of atomistic configurations consistent with the coarse-grained C$\alpha$ trace is not well represented. In contrast, DiAMoNDBack (PDB) -- 0.037 (PDB) and 0.033 (DES) -- and DiAMoNDBack (PDB;DES-FT) -- 0.064 (PDB) and 0.022 (DES) -- achieve diversity scores very close to the ideal value of zero, indicating that the distribution of configurational diversity between the backmapped atomistic configurations matches that of the generated configurations around the reference ground truth. The high bond scores and low clash scores for the DiAMoNDBack models indicate that despite the high configurational diversity, all of these various configurations are physically realistic with well-formed bonds and few steric collisions. The small performance boost in the DES test set using the DES fine-tuned model is indicative of a slight improvement in diversity resulting from the additional within-sample training, but the effect is almost negligible. Similarly, the small, but nearly negligible, performance degradation on the PDB test set indicates that the DES fine-tuned model is not overfitting. 

\begin{figure}[ht!]
\centering
\includegraphics[width=\columnwidth]{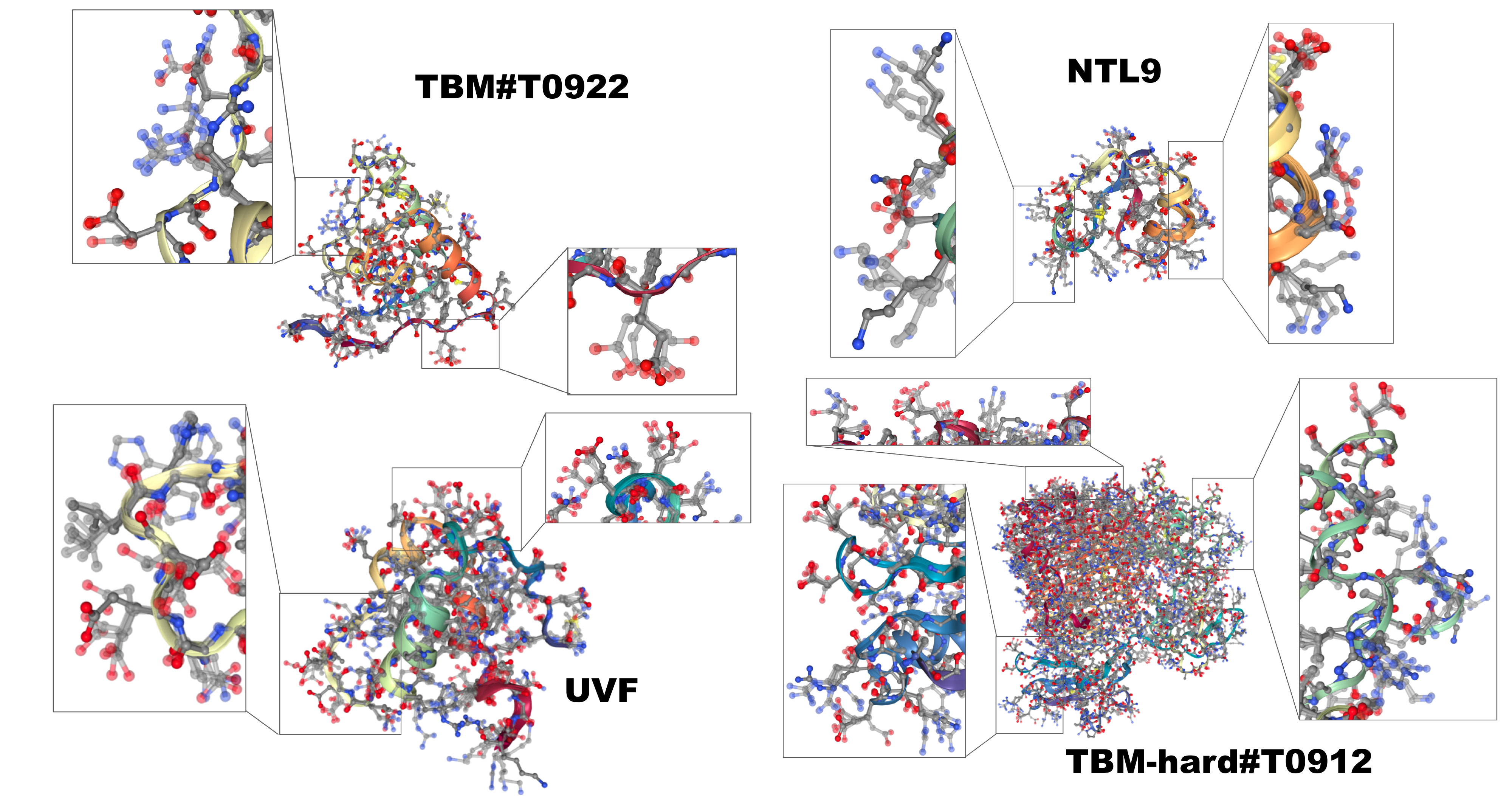}
\caption{Illustrative visualization of five DiAMoNDBack (PDB) atomistic backmappings for the best and worst performing examples from the PDB test set (best: TBM\#T0922; worst: TBM-hard\#T0912) and the DES test set (best: NTL9; worst: UVF) according to clash score. The five backmapped structures conditioned on the coarse-grained C$\alpha$ trace are shown as translucent structures and the single atomistic reference structure shown as opaque. Insets for each protein zoom into particular regions to highlight the generative diversity in side chain placements.} 
\label{fig:fig3}
\end{figure}

We illustrate the diversity of generated all-atom configurations in Fig.~\ref{fig:fig3} where we visualize five randomly generated backmappings for the best and worst-performing PDB and DES test set sequences reported in Table~\ref{tab:PDBDES}. As anticipated, we tend to see greater diversity of side chain configurations on the solvent-exposed exterior of the protein relative to those in the more tightly-packed hydrophobic core (\blauw{Fig.~S9-S10}) Interestingly, the N-to-C ordering of autoregressive decoding does not result in any detectable trends in the bond, clash, or diversity scores as a function of residue position in the protein chain (\blauw{Fig.~S8}). This suggests that the model has been well trained and achieves equally good backmapping quality irrespective of primary structure (i.e., sequence position) or location within the tertiary fold.

To further explore the impact of DES fine-tuning on the model performance we expose in Fig.~\ref{fig:fig4}a the distribution of bond and clash scores over five independently generated backmappings over the test set of 11 proteins in the DES data for the DiAMoNDBack (PDB) and DiAMoNDBack (PDB;DES-FT) models. The small but statistically significant improvement in the bond score upon fine tuning from (97.981$\pm$0.002)\% to (98.725$\pm$0.004)\% (Table~\ref{tab:PDBDES}) is visually apparent from the shift in probability mass in the violin plots towards 100\%. The clash score improvement from (0.33$\pm$0.01)\% to (0.175$\pm$0.007)\% (Table~\ref{tab:PDBDES}) is also statistically significant but less visually apparent in a shift in the per-sequence distributions due to the large population of frames with zero clashing residues. An average of $\sim$74\% frames across all sequences are generated with no clashes in all five samples for the PDB-trained model, which improves to $\sim$84\% frames with no clashes for the fine-tuned model. In Fig.~\ref{fig:fig4}b, we illustrate the improvement in model performance on each of the 11 proteins the DES test set by projecting the bond and clash scores of DiAMoNDBack (PDB) and DiAMoNDBack (PDB;DES-FT) into the plane. The migration of all points towards the upper-left of the plot after fine tuning indicates an across-the-board improvement in the bond and clash scores. An analogous analysis for the 24 proteins in the PDB test in Fig.~\ref{fig:fig4}c set shows a slightly different trend -- changes in bond and clash scores are mixed on the PDB test set after fine tuning on the DES training data, with a percent improvement of 8.77\% in clash score from (0.57$\pm$0.09)\% to (0.52$\pm$0.16)\% and a percent degradation of only 0.03\% in bond score from (99.18$\pm$0.04)\% to (98.99$\pm$0.06)\% (Table~\ref{tab:PDBDES}). These minor changes indicate that the fine-tuned model is not strongly overfit and the change in clash score actually lies within error bars. 

\begin{figure}[ht!]
\centering
\includegraphics[width=0.875\columnwidth]{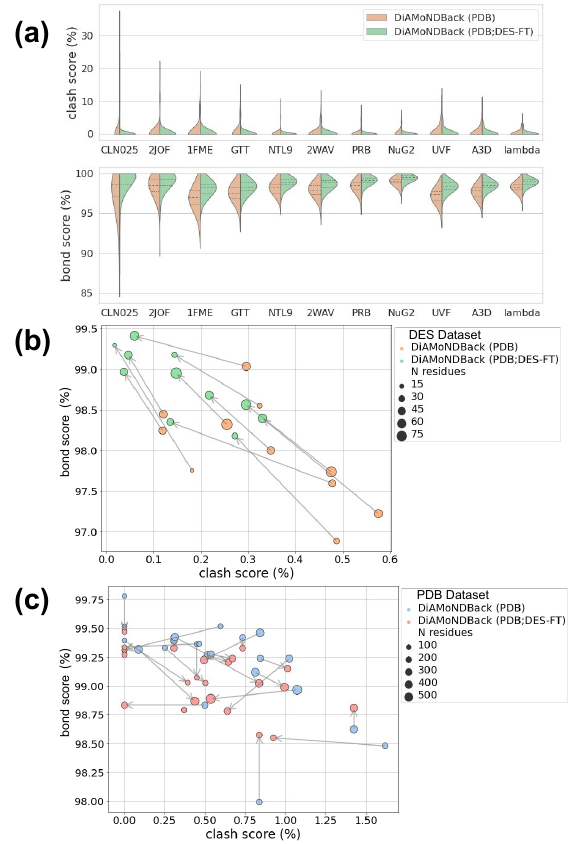}
\end{figure}

\begin{figure}[ht!]
\centering
\caption{Comparison of DiAMoNDBack (PDB) and DiAMoNDBack (PDB;DES-FT) models to assess the influence of DES fine-tuning upon performance. \textbf{(a)} Split violin plots of the distribution of clash scores (top) and bond scores (bottom) over five independently generated backmapped atomistic structures over the test set of 11 proteins in the DES data for the DiAMoNDBack (PDB) and DiAMoNDBack (PDB;DES-FT) models. As expected, both the bond scores and the clash scores improve with DES fine tuning. Scatter plots showing the change in bond score and clash score in applications of the DiAMoNDBack (PDB) and DiAMoNDBack (PDB;DES-FT) models to \textbf{(b)} the 11 proteins in the DES test set and \textbf{(c)} the 24 proteins in the PDB test set. Fine tuning on the DES training data results in an across-the-board improvement in bond and clash scores on the DES test set -- grey tie lines linking the results for the DiAMoNDBack (PDB) model applied to the DES test set (orange points) to those for the DiAMoNDBack (PDB;DES-FT) model applied to the DES test set (green points) illuminate a migration towards improved bond scores (up) and improved clash scores (left). Fine tuning results on the PDB test set show mixed results in the improvement/degradation of bond and clash scores -- grey tie lines link the results for the DiAMoNDBack (PDB) model applied to the PDB test set (blue points) to those for the DiAMoNDBack (PDB;DES-FT) model applied to the PDB test set (red points). Symbol size indicates the size of the test protein measured by number of residues. The lone red point that appears to not be unconnected to a blue point is due to the two points lying nearly on top of one another.}
\label{fig:fig4}
\end{figure}

Finally, comparing the performance of the baseline DiAMoNDBack (PDB) model, we observe generally better bond scores and poorer clash scores for the PDB test set compared to the DES test set. We can attribute the superior PDB bond scores to improved in-sample performance of the model fitted to the PDB training data. The inferior clash scores are more difficult to account for, but we suggest that they are likely a result of the PDB structures being much larger and more globular, while the MD simulation data represents smaller sequences with many frames each that undergo many folding transitions and spend time in more extended configurations that are less susceptible to clashes.

Taken together, this analysis shows that the DiAMoNDBack model trained over the PDB training data is capable of achieving competitive or superior accuracy in bond and clash scores to GenZProt and PULCHRA while also recapitulating a diverse ensemble of atomistic structures faithful to a particular C$\alpha$ coarse-graining. Fine tuning the model over the DES training data results in a slightly improved model for the DES test set without significant performance degradation over the PDB test set, and suggests a route to bespoke model training for particular molecular force fields.

\subsection{Analysis of side chain dihedral angles in generated atomistic structures} \label{sec:dih}

\begin{figure}[ht!]
\centering
\includegraphics[width=\columnwidth]{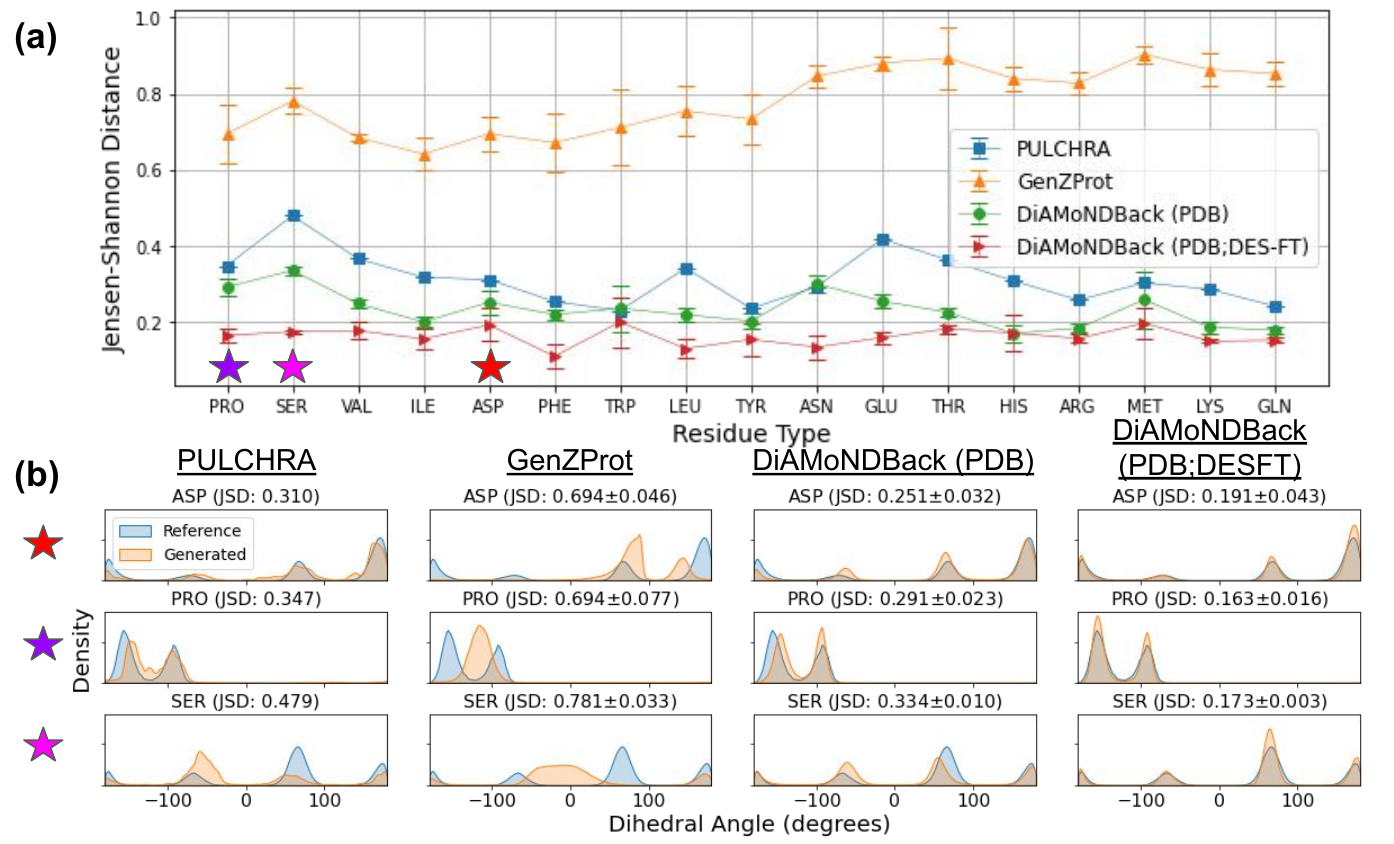}
\caption{Comparison of side chain C-C$\alpha$-C$\beta$-C$\gamma$ dihedral angles distributions over the test set simulation trajectories for the 11 DES proteins. \textbf{(a)} Jensen-Shannon divergences (JSD) are computed on a residue-wise basis between the molecular dynamics reference data and the backmapped configurational ensembles generated by PULCHRA, GenZProt, DiAMoNDBack (PDB), and DiAMoNDBack (PDB;DES-FT). Data are not reported for Gly and Ala, which do not possess this dihedral angle, and Cys, which is not represented in any of the 11 DES proteins. Error bars in the reported JSD values correspond to standard deviations and are estimated for the GenZProt and DiAMoNDBack models by averaging across five independent generations. The rules-based PULCHRA model is deterministic and has no associated standard deviation as all generations produced by this method are identical. \textbf{(b)} Comparison of the dihedral angle distributions for three selected residues Asp (red star), Pro (purple star), and Ser (pink star). Plots for all 17 residues are presented in \blauw{Fig.~S12}. The DiAMoNDBack (PDB) and DiAMoNDBack (PDB;DES-FT) models accurately recapitulate the multimodal dihedral angle distributions. The rules-based PULCHRA model tends to produce longer tails that overpopulate dihedral angles rarely visited in the reference data. The VAE-based GenZProt model encounters challenges in fitting the distributions possibly associated with mode collapse.} 
\label{fig:fig5}
\end{figure}

To further evaluate the structural fidelity of our backmapped structures we compared the distribution of side chain C-C$\alpha$-C$\beta$-C$\gamma$ dihedral angles of generated configurations relative to that collated from the test set simulation trajectories for the 11 DES proteins (Fig.~\ref{fig:fig5}). We conduct this analysis for 17/20 amino acids -- these dihedrals are not present in the small Gly and Ala side chains and none of the DES proteins contain Cys residues. Dihedral angle distributions for the generated configurations are calculated by generating five backmapped configurations for each frame of each protein in the DES test set molecular dynamics trajectories and collating normalized histograms of the dihedral angle distribution. An analogous procedure is used to compute the reference distribution directly from the test set simulation trajectories and we quantify the similarity of the two distributions using the Jensen-Shannon distance metric (JSD), which is the square root of the Jensen-Shannon divergence~\cite{lin1991divergence}. Employing a base two logarithm bounds the JSD to the range [0,1], with the lower bound of zero achieved for identical distributions. 

In Fig.~\ref{fig:fig5}a we compare the calculated JSD values for PULCHRA, GenZProt, DiAMoNDBack (PDB), and DiAMoNDBack (PDB;DES-FT). The DiAMoNDBack models exhibit superior performance for all amino acid residues with the fine-tuned model enjoying a small benefit in performance for all residues except His. We present a comparison of the dihedral angle distributions for three selected residues in Fig.~\ref{fig:fig5}b, with the remaining residues presented in \blauw{Fig.~S12}. The DiAMoNDBack (PDB) and DiAMoNDBack (PDB;DES-FT) models produce distributions in excellent agreement with the molecular dynamics reference and accurately recapitulate the multimodal nature of these distributions. The rules-based PULCHRA model is able to reproduce this multimodality but tends to produce longer tails and exhibits significantly poorer agreement to the reference distribution. PULCHRA operates by first generating an approximate structure using heuristics and atomic fragments followed by rotation of side chain dihedrals to resolve steric clashes~\cite{rotkiewicz2008fast}. We attribute the tails to this second step that overpopulates regions of side chain dihedral space that are unrepresented in the reference data. On average, we observe 26\% lower JSD scores compared to PULCHRA for DiAMoNDBack (PDB) and this improves to 49\% for DiAMoNDBack (PDB;DES-FT). GenZProt model exhibits the poorest agreement to the reference data and seemingly encounters challenges in mimicking the multimodal distributions that we hypothesize may be attributable to mode collapse within the GenZProt VAE~\cite{lucas2019understanding}. An analysis of internal energies in the backmapped configurations demonstrates that the DiAMoNDBack models also perform well in reproducing these distributions (\blauw{Fig.~S13-S14}), but it is important to exercise caution in interpreting these results for the purposes of a structural comparison due to the high sensitivity of energy potentials to minor structural changes. We also recall that \textit{post-hoc} energy minimization could be conducted to yield low-energy configurations but that we elect not to perform any refinement for the purposes of restricting our comparisons to the backmapping methodology alone without biasing to a particular molecular force-field and avoiding the large computational cost associated with these operations for large proteins.

\subsection{Analysis of residue-wise performance}

\begin{figure}[ht!]
\centering
\includegraphics[width=0.98\columnwidth]{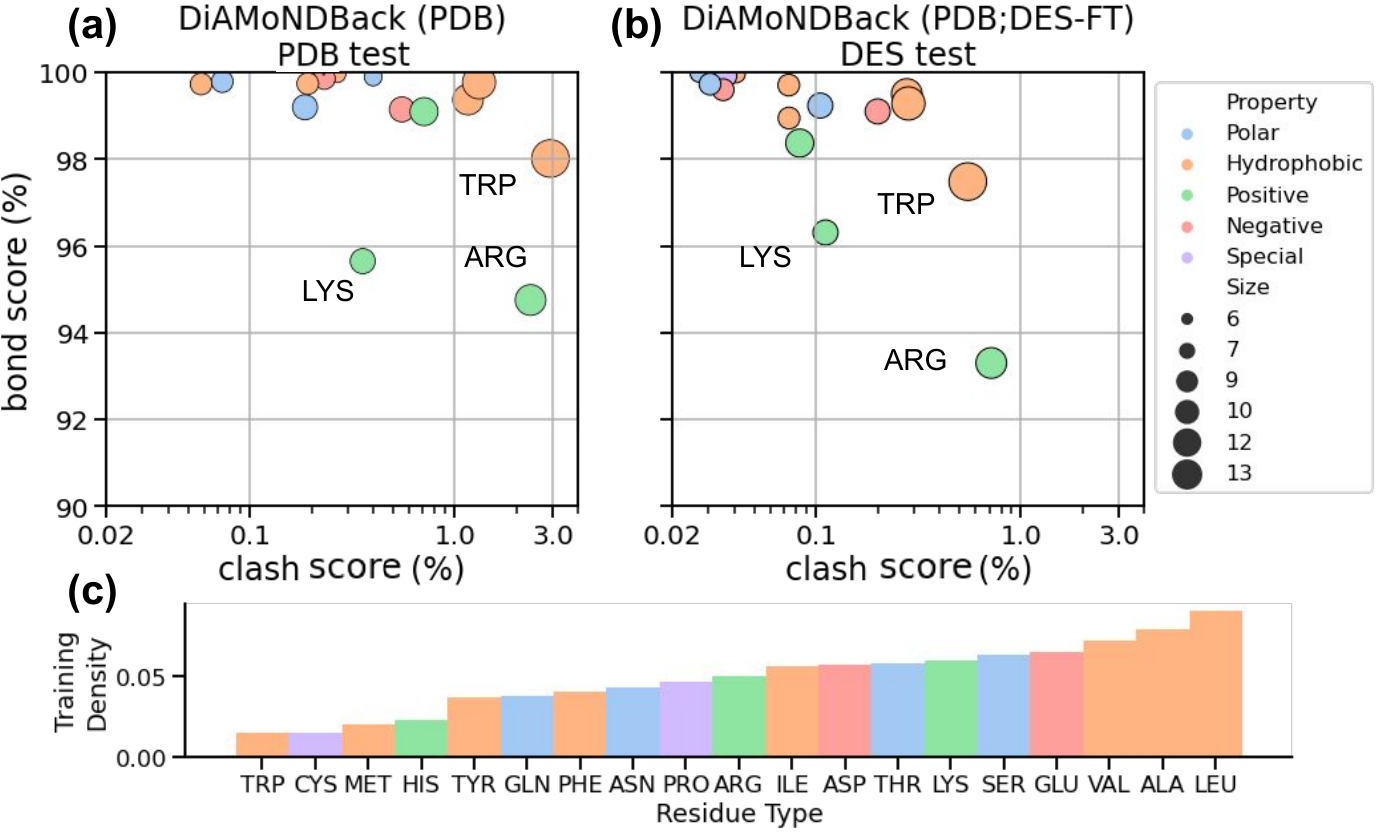}
\caption{Analysis of DiAMoNDBack residue-wise bond and clash scores. Scatterplots illustrating the residue-wise bond and clash scores for \textbf{(a)} the DiAMoNDBack (PDB) model evaluated on the PDB test set and \textbf{(b)} the DiAMoNDBack (PDB;DES-FT) model evaluated on the DES test set. Markers are sized according to the number of atoms in the corresponding side chain and colored by phsyicochemical grouping: polar, hydrophobic, positively charged, negatively charged, and special (Pro and Cys). \textbf{(c)} Probability distribution illustrating representation of each residue type within the PDB training data. The bar heights are normalized to sum to unity.}
\label{fig:fig6}
\end{figure}

We next sought to explore whether particular residue types were more prone to produce poorly formed bonds and be involved in unphysical steric collisions. Identifying residue-level trends in these performance metrics can help expose potential failure modes for our model. In Fig.~\ref{fig:fig6} we present scatter plots of the bond and clash scores broken down on a per residue basis. Application of DiAMoNDBack (PDB) to the PDB test set shows that all residues possess clash scores of 3\% or less, and all but three residues -- Arg, Lys, and Trp -- possess bond scores better than 98\% (Fig.~\ref{fig:fig6}a). This trend is maintained in application of DiAMoNDBack (PDB;DES-FT) to the DES test set, with all residues possessing good clash scores of 1\% or better, and all but Arg, Lys, and Trp possessing bond scores better than 98\%. The three outliers in each case still possess good bond scores better than 93\%, but it is informative to understand this relatively poorer behavior. Arg and Lys both possess long, charged side chains that are exceptionally dynamic, can rapidly exchange their protons with water, and are challenging to resolve experimentally~\cite{nguyen2019nmr, esadze2011dynamics}. This observation is consistent with our observation during our data cleaning procedure that Arg and Lys residues tended to possess significantly more incomplete side chains within the PDB data set compared to other residues, potentially indicating lower confidence in experimental resolution of side chain atomic coordinates. Trp is both the bulkiest amino acid, and therefore potentially the most susceptible to steric clashes. It is also the least represented within the PDB training data (Fig.~\ref{fig:fig6}c) with the relatively smaller number of training examples meaning that the model may be less well trained on this residues and less able to generalize to unfamiliar configurational environments.

\subsection{Evaluation on CG trajectories}

To illustrate a realistic application of the backmapping models to coarse grained molecular trajectories, we test the capability of PULCHRA, GenZProt, and DiAMoNDBack to restore atomistic detail to simulation trajectories of three small proteins -- 1FME (28 residues), PRB (47 residues), and A3D (73 residues) -- generated by Majewski et al.\ using bespoke C$\alpha$-based coarse-grained potentials~\cite{majewski2022machine} (Table~\ref{tab:CG}). Compared to our prior analyses, these data represent an out-of-distribution test case generated by a coarse-grained force field to which the model was never exposed during training and for which there is no atomistic ground truth. The bond scores for all four models are excellent and on par with those for the in-sample PDB and DES tests reported in Table~\ref{tab:PDBDES}. The clash scores are slightly poorer but for PULCHRA and DiAMoNDBack are still very good, achieving 2\% or fewer clashes for all three proteins. The GenZProt clash score is quite poor at 8-15\%. Without atomistic structures we cannot use Eqn.~\ref{eqn:div} to compute the diversity score, so instead compute $RMSD_{gen}$ using Eqn.~\ref{eqn:gen} as a proxy for generative diversity. The deterministic PULCHRA model generates zero diversity, while GenZProt, DiAMoNDBack (PDB), and DiAMoNDBack (PDB;DES-FT) generate $RMSD_{gen}$ values of, respectively, 0.212 nm, 1.69 nm, and 1.57 nm on average. The baseline DiAMoNDBack (PDB) model produces an average $RMSD_{gen}$ increase over GenZProt of nearly 8$\times$ across these three sequences. Finally, in Fig.~\ref{fig:fig7} we present illustrative visualizations of the atomistic backmappings for these three coarse-grained proteins generated by DiAMoNDBack (PDB). These results illustrate that DiAMoNDBack is capable of performing physically realistic and diverse atomistic backmappings for out-of-distribution coarse grained simulation trajectories produced by a C$\alpha$-based coarse-grained model.

\begin{table}[]
\resizebox{\columnwidth}{!}{%
\begin{tabular}{@{}cccc@{}}
\cmidrule(l){2-4}
            & CG 1FME         & CG PRB           & CG A3D           \\ \cmidrule(l){2-4} 
            & Bond ($\uparrow$) {[}\%{]} & Bond ($\uparrow$) {[}\%{]}  & Bond ($\uparrow$) {[}\%{]}  \\ \cmidrule(l){2-4} 
PULCHRA & 98.93           & 99.18            & 99.60            \\
GenZProt    & 95.57$\pm$0.01  & 97.210$\pm$0.003 & 96.930$\pm$0.001 \\
DiAMoNDBack (PDB)  & 96.72$\pm$0.01  & 98.23$\pm$0.01   & 98.25$\pm$0.01   \\
DiAMoNDBack (PDB;DES-FT) & 97.81$\pm$0.03 & 98.97$\pm$0.02 & 98.64$\pm$0.01 \\ \cmidrule(l){2-4} 
            & Clash ($\downarrow$) {[}\%{]} & Clash ($\downarrow$) {[}\%{]}  & Clash ($\downarrow$) {[}\%{]}  \\ \cmidrule(l){2-4} 
PULCHRA & 1.29            & 0.673            & 0.298            \\
GenZProt    & 14.01$\pm$0.03  & 8.20$\pm$0.05    & 11.88$\pm$0.01   \\
DiAMoNDBack (PDB)  & 1.96$\pm$0.04   & 1.12$\pm$0.05    & 1.06$\pm$0.04    \\
DiAMoNDBack (PDB;DES-FT) & 1.75$\pm$0.06  & 0.97$\pm$0.03  & 1.08$\pm$0.04  \\ \cmidrule(l){2-4} 
            & RMSD ($\uparrow$) {[}nm{]}  & RMSD ($\uparrow$) {[}nm{]}   & RMSD ($\uparrow$) {[}nm{]}   \\ \cmidrule(l){2-4} 
PULCHRA & 0              & 0               & 0               \\
GenZProt    & 0.23$\pm$0.09   & 0.225$\pm$0.070  & 0.18$\pm$0.02    \\
DiAMoNDBack (PDB)  & 1.96$\pm$0.29   & 1.43$\pm$0.16    & 1.67$\pm$0.17    \\
DiAMoNDBack (PDB;DES-FT) & 1.80$\pm$0.27  & 1.30$\pm$0.14  & 1.57 $\pm$0.15 \\ \cmidrule(l){2-4} 
\end{tabular}%
}
\caption{Application of backmapping to bespoke coarse-grained trajectories generated by Majewski et al.~\cite{majewski2022machine}. Comparisons are presented for three sequences spanning a range of lengths: 1FME (28 residues), PRB (47 residues), and A3D (73 residues). Standard deviations in the bond and clash scores reported for DiAMoNDBack and GenZProt are estimated using five-fold block averaging for the bond and clash scores. As a deterministic algorithm, the values reported for PULCHRA do not have associated uncertainties. In the absence of an atomistic ground truth, the average pairwise RMSD between generated samples is reported as a proxy for generative diversity.}
\label{tab:CG}
\end{table}

\begin{figure}[ht!]
\centering
\includegraphics[width=0.6\columnwidth]{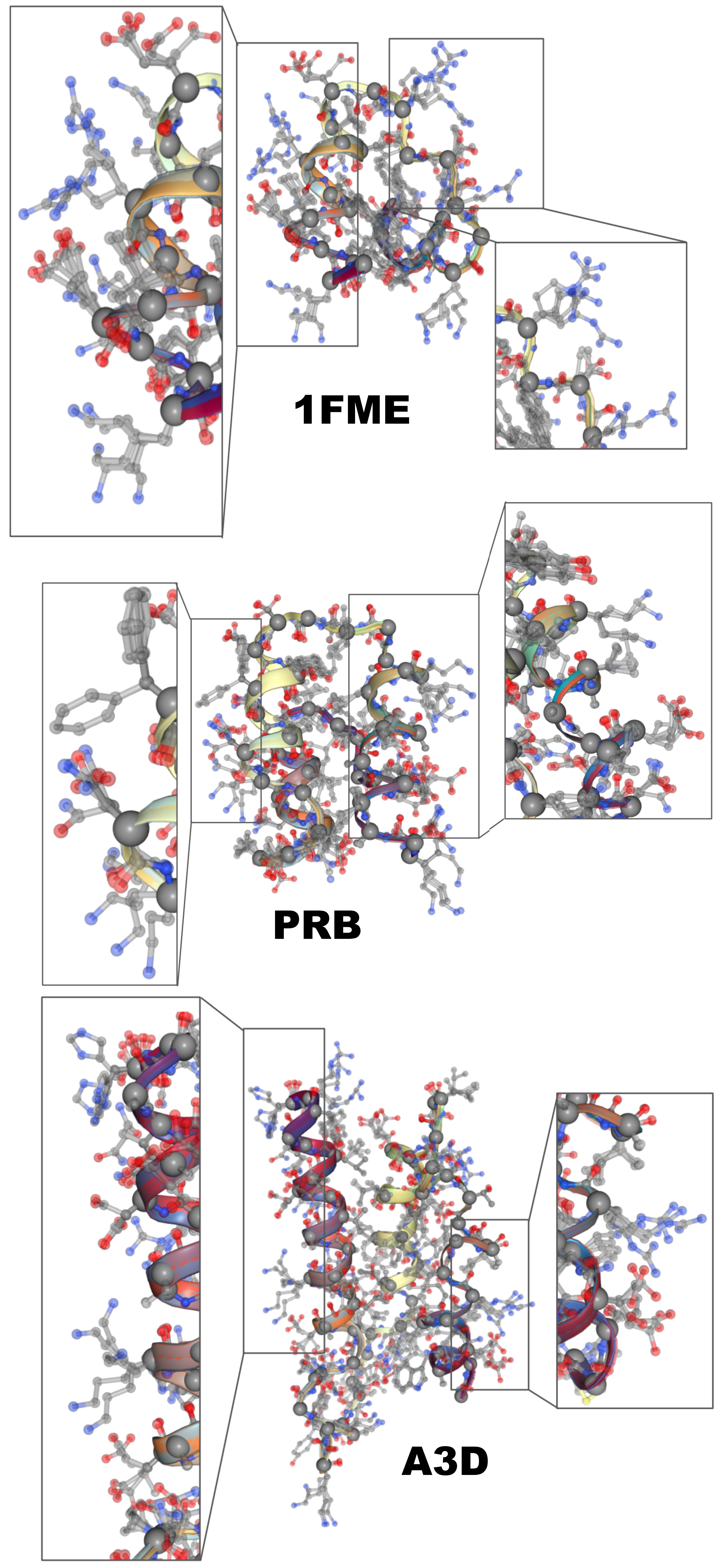}
\end{figure}

\begin{figure}[ht!]
\centering
\caption{Illustrative DiAMoNDBack (PDB) atomistic backmappings for coarse-grained simulation trajectories of the 1FME (28 residues), PRB (47 residues), and A3D (73 residues) proteins generated by Majewski et al.\ using bespoke C$\alpha$-based coarse-grained potentials~\cite{majewski2022machine}. Five indepnedent backmapped structures conditioned on the coarse-grained C$\alpha$ trace are shown as translucent and the single C$\alpha$ trace shown as opaque. Insets for each protein zoom into particular regions to highlight the generative diversity in side chain placements.} 
\label{fig:fig7}
\end{figure}

\section{Conclusions}

In this work, we present DiAMoNDBack (Diffusion-denoising Autoregressive Model for Non-Deterministic Backmapping) as a transferable approach for backmapping C$\alpha$ protein traces into atomistic structures using Denoising Diffusion Probabilistic Models (DDPMs). Our approach builds the protein structure in an autoregressive manner residue-by-residue, at each instance predicting the Cartesian coordinates of the target residue aligned to a canonical reference frame conditioned on all previously decoded residues within a local neighborhood and the coarse-grained C$\alpha$ trace. We train models on a corpus of 65k+ PDB structures exposing our model to a rich variety of local residue environments to establish a general-purpose model for backmapping generic C$\alpha$ traces. Evaluating our approach on held-out PDB structures and all-atom molecular dynamics simulations reveals excellent performance in terms of the quality of the reconstructed atomic bonds, the degree of non-bonded steric clashes between residues, and the capacity to generate a diverse ensemble of atomistic structures consistent with a particular C$\alpha$ trace. While we find improved structural metrics compared to the previous transferable backmapping work of Yang and G\'omez-Bombarelli~\cite{yang2023chemically} (GenZProt) and are competitive with rules-based approaches~\cite{rotkiewicz2008fast} (PULCHRA), the generative diversity is where our model excels in producing a substantially more diverse ensemble of atomistic reconstructions consistent with the coarse-grained C$\alpha$ trace. Analysis of side chain dihedral angles in reconstructed structures also reveal our model generates more physically plausible distributions of internal coordinates compared to the rules-based approach that involves manually adjusting side chain dihedral angles to resolve clashes. We demonstrate a deployment of our model to coarse-grained simulation trajectories generated by bespoke C$\alpha$-based coarse-grained force fields, and show that it can generate high-quality bonds ($\sim$98\% bond lengths within 10\% of reference data) and low fractions of clashing residues ($\sim$1.25\%). We also demonstrate fine-tuning of our baseline PDB-trained DiAMoNDBack model on limited numbers of all-atom simulation data to develop a force field-specific model with slightly improved performance on those data. We believe that DiAMoNDBack offers a generic, transferable, and accurate backmapping tool of value to the community and we have made it freely available as an open source Python package (see \blauw{Data Availability Statement}).

In future work we would like to investigate a number of innovations to further enhance the quality of our atomistic reconstructions. Our model currently generates atomistic residues from C$\alpha$ traces conditioned on the $N$-nearest neighboring residues. Our current data representation exposes the Cartesian coordinates of the local environment and imposes practical limits on the conditioning size due to increasing dimensionality slowing training and inference efficiencies. Residues outside the purview of this local neighborhood can therefore be excluded from the conditioning effectively hidden from the model and potentially leading to clashes. A more comprehensive conditioning scheme that incorporates the full protein structure, for instance by using a graph neural network to assemble the conditioning information, could potentially improve the quality of our generated structures. One significant drawback of our model is that the DDPM generation process is relatively slow, making DiAMoNDBack approximately 50$\times$ slower than GenZProt and 100$\times$ slower than PULCHRA (\blauw{Fig.~S7}). Speed-ups could be achieved by treating non-interacting regions of a structure independently and decoding residues in parallel when possible to accelerate upon sequential N-to-C decoding. While in this work we focus on backmapping from C$\alpha$ traces, we observe that the framework is extensible to any resolution of coarse-graining and could readily be applied to multi-site per residue models such as MARTINI~\cite{marrink2007martini, monticelli2008martini, souza2021martini} and AWSEM~\cite{davtyan2012awsem} or even multi-residue per site ultra coarse-grained models~\cite{dama2013theory, trylska2010coarse}. Furthermore, growing repositories of nucleic acid protein complexes~\cite{sagendorf2020dnaprodb} can be used to train a model that backmaps from DNA-protein coarse-grained forcefields such as AWSEM-3SPN.2~\cite{lu2021openawsem} or GENESIS-CG~\cite{tan2022implementation}.


%

\section*{Data and Code Availability}
We make our backmapping model publicly available by releasing pre-trained models and code for use at \url{https://github.com/Ferg-Lab/DiAMoNDBack}. We also make available all the data splits used to train and test the models reported in this work available via Zenodo at \href{https://doi.org/10.5281/zenodo.8169238}{DOI:10.5281/zenodo.8169238}~\cite{zenodo}.

\begin{acknowledgement}
This material is based on work supported by the National Science Foundation under Grant No.\ CHE-2152521. K.S.\ was supported by a fellowship from the Molecular Sciences Software Institute under the National Science Foundation Grant No.\ CHE-2136142. This work was completed in part with resources provided by the University of Chicago Research Computing Center. We gratefully acknowledge computing time on the University of Chicago high-performance GPU-based cyberinfrastructure supported by the National Science Foundation under Grant No.\ DMR-1828629. We thank Adri\`a P\'erez for their help in sharing the coarse-grained trajectories. We are grateful to D.E. Shaw Research for sharing the protein simulation trajectories.
\end{acknowledgement}

\begin{suppinfo}
Additional information on the mathematical formalism, DDPM loss function, conditional U-net architecture, residue featurization, canonical alignment process, treatment of N- and C-terminal residues, model hyperparameter tuning, residue representation in our training data, usage of the PULCHRA and GenZProt models, analysis of reconstructed terminal residue quality, dihedral angle distributions for all residues, error as a function of residue position along the protein chain, and analysis of internal energies.
\end{suppinfo}

\section*{Conflict of Interest Statement}

A.L.F.\ is a co-founder and consultant of Evozyne, Inc.\ and a co-author of US Patent Applications 16/887,710 and 17/642,582, US Provisional Patent Applications 62/853,919, 62/900,420, 63/314,898, 63/479,378, and 63/521,617, and International Patent Applications PCT/US2020/035206 and PCT/US2020/050466.


\clearpage
\newpage

\bibliography{ref}


\clearpage
\newpage

\begin{tocentry}


%
%
\begin{center}
\includegraphics[width=2.5in]{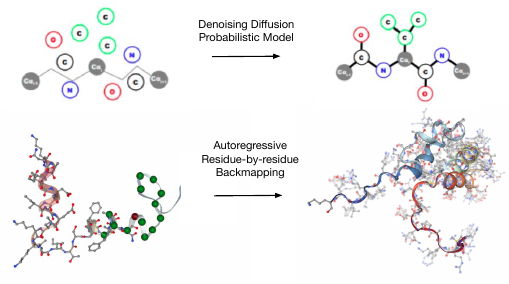}
\end{center}
\end{tocentry}

\end{document}